\documentclass[12pt]{article}

\usepackage{verbatim}
\usepackage{hyperref}
\usepackage{color}
\usepackage{amsfonts}
\usepackage{amsmath}
\usepackage{amssymb}
\usepackage{latexsym}
\usepackage{lastpage}
\usepackage[pdftex]{graphicx} 
\usepackage[center]{caption2}

\date{}
\begin{document}
\begin{flushright}
\hfill UPR-1283-T\
\end{flushright}

\vspace{20pt}
\begin{center}
{\Large {\bf Vacuum polarization throughout general subtracted black hole spacetimes}}

\vspace{18pt}
{\large   Mirjam Cveti\v c$^{\dagger\,\ddagger}$ and Alejandro Satz$^{\dagger\dagger}$}

\vspace{8pt}

{\it  $\dagger$ Department of Physics and Astronomy,  University of Pennsylvania, Philadelphia, PA 19104, USA\\ $\ddagger$  Center for Applied Mathematics and Theoretical Physics, University of Maribor, SI2000 Maribor, Slovenia 

\vspace{5pt}
 
{\it  $\dagger\dagger$ }Institute for Gravitation and the Cosmos \& Physics Department, Penn State, University Park, PA 16802, USA}

\end{center}
\vspace{26pt}

\begin{abstract}
We compute the vacuum polarization of a massless minimally coupled scalar field in a background given by a black hole with subtracted geometry. Extending previous results for the horizon of rotating black holes with no charge, we obtain an analytical expression for the vacuum polarization that is valid throughout the spacetime and for arbitrary rotation and charge parameters. The vacuum polarization diverges at the inner horizon and the quantum state cannot be extended to the inside of it. 
\end{abstract}

\section{Introduction}
Quantum field theory in curved spacetime is a semiclassical approximation to quantum gravity that describes the behavior of quantum fields in gravitational backgrounds. It has a wide range of physically important applications, notably in inflationary cosmology \cite{Mukhanov:1981xt} and in black hole evaporation \cite{Hawking:1974sw}. Since the particle concept is usually ambiguous or inapplicable on curved backgrounds, attention is often given to local covariant observables such as the vacuum polarization $\langle\phi^2\rangle$ and  the stress energy tensor $\langle T_{\mu\nu}\rangle$. The second of these has more direct physical relevance as the source for the backreaction of the matter fields upon spacetime. The vacuum polarization, however, is important as a preliminary step for the computation of  $\langle T_{\mu\nu}\rangle$, as well as on its own right as a direct scalar probe of quantum fluctuations and  as affecting e.g. symmetry breaking computations.

There is a long history to the attempts to compute vacuum polarization in black hole backgrounds, starting with Candelas' evaluation at the horizon of a Schwarzschild black hole \cite{Candelas:1980zt}. For asymptotically flat static black hole solutions, Candelas' methods cannot be extended beyond the horizons but techniques for numerical evaluation have been developed \cite{Anderson:1990jh}. For the Kerr and Kerr-Newman black holes analytical results are only available at the horizon pole \cite{Frolov:1982pi}. Numerical evaluations throughout the horizon were first obtained in \cite{Duffy:2005mz}, and recently a method for numerical evaluation throughout the spacetime was outlined in \cite{Levi:2016exv,Levi:2016paz}. Analytic results throughout the spacetime are available in three dimensions and with AdS asymptotics \cite{Krishnan,Louko} but are generally impossible in four dimensions, where even with AdS asymptotics numerical evaluations are required \cite{Flachi:2008sr}.

In this paper we will show that an exact expression for the vacuum polarization $\langle\phi^2\rangle$ of a massless, minimally coupled scalar field is in fact obtainable, throughout the black hole spacetime, for the class of black holes known as subtracted geometry.  Subtracted geometry black holes  \cite{Cvetic:2011hp,Cvetic:2011dn,Cvetic:2012tr,Cvetic:2014sxa} are  solutions of the bosonic sector of N=2 STU supergravity coupled to three vector multiplets. (The general asymptotically flat black holes of the STU model were constructed in \cite{Cvetic:1996kv,Chow:2013tia,Chow:2014cca}.) The subtracted black hole metric can be obtained by subtracting some terms in the ``warp factor'' of the original black hole metric in such a way that the massless minimally coupled scalar  wave equation  becomes separable and analytical solutions are obtainable. This subtracted black hole metric effectively places the black hole in an asymptotically conical box and mimics the ``hidden conformal symmetry'' 
\cite{Castro:2010fd}
 of the wave equation on rotating black holes in the near-horizon, near-extremal, and/or low energy regimes, which is a key motivator for the Kerr/CFT conjecture (see e.g. 
 \cite{Compere:2012jk}). The energy density of the matter fields in this new geometry falls off as second power of radial distance, thus confining thermal radiation. The classical near-horizon properties of the subtracted black hole are the same as the original black hole ones; in particular, the classical thermodynamics of the subtracted black hole is analogous to the standard one 
 \cite{Cvetic:2014nta},
although loop corrections to the horizon entropy differ 
\cite{Cvetic:2014tka}). 

The horizon vacuum polarization was studied for static subtracted black holes in \cite{Cvetic:2014eka}
and for rotating uncharged subtracted black holes in \cite{Cvetic:2015cca}. In this paper we extend the results of \cite{Cvetic:2015cca} in a twofold way. Firstly, we allow for general values of the four charge parameters associated to the subtracted metric, in addition to angular momentum, thus considering the most general possible subtracted black hole solution. Secondly and most importantly, we compute the vacuum polarization throughout the spacetime, both outside and inside the horizon, instead of just at the horizon. This method we use is to compute first the Feynman  Green's function of the massless scalar on this background and then take the coincidence limit, adding suitable counterterms to cancel the arising divergences. The Green's function is in turn computed by dimensional reduction from the five-dimensional $AdS^3\times S^2$ spacetime, in which the subtracted geometry can be embedded. As we will show, the quantum state defined by this procedure is the subtracted geometry version of the Hartle-Hawking thermal vacuum.

This paper is organized as follows. In the Section 2, we  first introduce the subtracted black hole metric and its five-dimensional embedding, and then discuss the Green's function on this background. In Section 3 we take the regularized coincidence limit in the Green's function, obtaining an expression for the vacuum polarization that is our main result. In Section 4 we  discuss particular cases and limits of this expression, and in the final concluding section we  summarize the results and discuss prospects for future work.

\section{Green's function on general subtracted black hole background}

\subsection{The subtracted geometry}

The general four-dimensional axisymmetric black hole metric is given by:
\begin{equation}
d  s_4^2  = -  \Delta^{-1/2}  G \, 
( d{ t}+{ {\cal  A\, \mathrm{d}\varphi}})^2 + { \Delta}^{1/2}
\left(\frac{d r^2} { X} + 
d\theta^2 + \frac{ X}{  G} \sin^2\theta\, d\varphi^2 \right)\,.\label{metric4d}
\end{equation}
The quantities $X, G, {\cal A},  \Delta $ are all functions of $r$ and $\sin \theta$ only (and depend on  the mass, rotation and charge parameters). For a given conventional (asymptotically flat) black hole solution with mass $M$, angular momentum $J$, and up to four charge parameters $Q_I$, we can construct a corresponding subtracted black hole solution by modifying only the so-called warp factor $\Delta(r,\theta)$. Specifically, a subtracted black hole geometry is given by:
\begin{eqnarray}\label{subfunctions}
{ X} & =& { r}^2 - 2{ m}{ r} + { a}^2~,\cr
{ G} & = &{ r}^2 - 2{ m}{ r} + { a}^2 \cos^2\theta\,, \cr
{ {\cal A}}  &=&{2{ m} { a}{G^{-1}} \sin^2\theta }
\left[ ({ \Pi_c} - { \Pi_s}){  r} + 2{ m}{ \Pi}_s\right]\,,\cr
\Delta &=& (2 m)^3 r ( \Pi_c^2 -\Pi_s^2) + (2m)^4 \Pi_s^2 - (2m)^2 ( \Pi_c-\Pi_s)^2 a^2 \cos^2 \theta\,,
  \label{c4d}
\end{eqnarray}
where the black hole parameters are encoded as:
\begin{eqnarray}
Q_I & = &{\frac{1}{4}}m \sinh 2\delta_I~,~(I=0,1,2,3)~,\nonumber\\
M  &= &{\frac{1}{4}}m\sum_{I=0}^3\cosh 2\delta_I ~, \label{Mdef} \nonumber\\
J & = &m\, a \,(\Pi_c - \Pi_s)~,~\Pi_c = \prod_{I=0}^3\cosh\delta_I 
~,~\Pi_s =  \prod_{I=0}^3 \sinh\delta_I~.
\end{eqnarray}
In a conventional black hole the warp factor is a fourth-order polynomial in $r$; for example, the Reissner-Nordstrom metric is obtained setting $a=0$,~$\delta_I=\delta$ in $X,G,\cal A$ above and $\Delta=(r+2m\sinh\delta)^4$, whereas the subtracted Reissner-Nordstrom metric has $\Delta = (2m)^3[r(  \Pi_c^2 -\Pi_s^2) +2m\Pi_s^2]$ instead. 
In both the original and the subtracted case the horizons, specified by $X=0$, are at:
\begin{equation}\label{defhorizons}
r_\pm=m\pm \sqrt{m^2-a^2}\, .
\end{equation}

Subtracted black holes are solutions of the bosonic sector four-dimensional ${\cal N}=2$ supergravity coupled to three vector supermultiplets; the detailed form of matter fields supporting the geometry is given in \cite{Cvetic:2011dn,Cvetic:2012tr}. An important feature of subtracted black holes is that they have a natural embedding in $AdS^3\times S^2$ \cite{Cvetic:2013lfa}. The BTZ black hole metric can be written as:
\begin{align}
ds_{BTZ}^2=&-\frac{(R^2-_+^2)(R^2-R_-^2)}{l^2R^2}dt_3^2+\frac{l^2R^2}{(R^2-R_+^2)(R^2-R_-^2)}dR^2\nonumber\\
&+R^2\left(d\varphi_3+\frac{R_+R_-}{l R^2}dt_3\right)^2\,,
\end{align}
which is locally isometric to $AdS^3$ with radius $l$. Consider the $5d$ manifold with metric
\begin{equation}
ds_5^2=ds_{BTZ}^2+ds_{S^2}^2\,,\quad ds_{S^2}^2=\frac{l^2}{4}\left(d\theta^2+\sin^2\theta\, d\bar{\varphi}^2\right)
\end{equation}
To obtain the 4d subtracted black hole metric in the $(t,r,\theta,\varphi)$ coordinates as given above in (\ref{metric4d}), we first need to make the identifications:
\begin{eqnarray}\label{identifications}
\bar{\varphi} &=& \varphi-\frac{16m a(\Pi_c-\Pi_s)}{l^3}(z+t)\cr
R^2&=&\frac{64m^2R_0^2}{l^4}\left[2mr(\Pi_c^2-\Pi_s^2)+4m^2\Pi_s^2-a^2(\Pi_c-\Pi_s)^2\right]\cr
R_\pm&=&\frac{8mR_0}{l^2}\left[m(\Pi_c+\Pi_s)\pm\sqrt{m^2-a^2}(\Pi_c-\Pi_s)\right]\cr
\varphi_3&=&z/R_0\,;\quad\quad t_3=(l/R_0)t\,;\quad\quad l=4m(\Pi_c^2-\Pi_s^2)^{1/3}
\end{eqnarray}
where $R_0$ is an arbitrary lengthscale. Then we can write:
\begin{equation}
ds_5^2=\frac{Q^2}{\sqrt{\Delta}}ds_4^2+\frac{\Delta}{Q^4}\left(dz+\mathcal{A}_2\right)^2\,,
\end{equation}
where $\Delta$ is given above in (\ref{subfunctions}), and:
\begin{eqnarray}
\mathcal{A}_2 & = & \frac{Q^3[4m^2\Pi_c\Pi_s+a^2(\Pi_c-\Pi_s)^2\cos^2\theta]}{2m(\Pi_c^2-\Pi_s^2)\Delta}dt+\frac{Q^32ma(\Pi_c-\Pi_s)\sin^2\theta}{\Delta}d\varphi\, \cr
Q &=& 2m(\Pi_c^2-\Pi_s^2)^{1/3}=l/2 .
\end{eqnarray}

\subsection{The Green's function}

The Green's function of a massless, minimally coupled scalar field on $AdS^3\times S^2$ (with radii $l$,$l_2$ respectively) has been computed in \cite{Bena:2010gg}. It takes the form:
\begin{equation}
G_5(x,x')=\frac{1}{8\sqrt{2}\pi^2 l \,l_2^2}\frac{\zeta}{[2\zeta^2-1-\cos\gamma]^{3/2}}\,,
\end{equation}
where $\zeta=\zeta(x,x')$ and $\cos\gamma =\cos\gamma(x,x')$ are related to the $AdS^3$ and the $S^2$ distances respectively:
\begin{eqnarray}\label{zetagamma}
\zeta(x,x') &=& \frac{\Delta X^2}{2l^2}+1 \cr
\cos\gamma(x,x') &=& \cos\theta\cos\theta'+\sin\theta\sin\theta'\cos(\bar{\varphi} -\bar{\varphi} ')
\end{eqnarray}
where $\Delta X^2(x,x')$ is the distance in the Minkowski space with signature $(--++)$ where $AdS^3$ is embedded.

The Green's function on the subtracted black hole background can be obtained by setting $l=2l_2$, expressing $G_5(x,x')$ in the $(t,r,\theta,\varphi,z)$ coordinates, and then integrating over the embedding coordinate $z$:\footnote{This works because $\partial_z$ is a Killing vector of the 5d manifold, and because the extra term $\mathcal{A}_2$ in the metric is a one-form Kaluza-Klein gauge potential and thus the transformation $dz\rightarrow dz+\mathcal{A}_2$ when integrating does not alter the Jacobian.} 
\begin{equation}\label{greenz}
G_4(t,r,\theta,\varphi;t'r',\theta',\varphi')  =\frac{1}{2\sqrt{2}\pi^2l^3}\int_{-\infty}^{+\infty}dz\frac{\zeta(z;0)}{[2\zeta^2(z;0)-1-\cos\gamma(z;0)]^{3/2}}
\end{equation}
Here the dependence of $\zeta$ and $\cos\gamma$ on the eight coordinates that $G_4$ depends on is kept implicit. The full form of $\zeta(t,r,z;t'r',0)$  differs depending on which of the six possible combinations of the three ranges of the black hole radial coordinate $(0,r_-), (r_-,r_+), (r_+,+\infty)$ is the one where $(r,r')$ fall into. For example, in the external region where both $r,r'>r_+$ and for $t=t'$, we have
\begin{align}\label{zetar}
\zeta(t,r,z;t,r',0) =& \frac{1}{r_+ - r_-}\Big[[\sqrt{(r-r_-)(r'-r_-)}\cosh(c_+ z)\nonumber\\
&-\sqrt{(r-r_+)(r'-r_+)}\cosh(c_- z)\Big]
\end{align}
 where 
 \begin{equation}\label{alphadef}
c_\pm = \frac{16m^2}{2(1+\alpha^2)l^3}\left[(1+\alpha^2)(\Pi_c+\Pi_s)\pm (1-\alpha^2)(\Pi_c-\Pi_s)\right]\,,\quad\alpha\equiv a/r_+
\end{equation}
For the other six possible combinations of radial ranges, (\ref{zetar}) gets sign changes at the square roots, as well as cosh functions changed to sinh when the points fall in different ranges. The exact form of $\zeta$ for each range of the two radial coordinates is given in the Appendix.  $\alpha$ will be instead of $a$ as a more convenient rotation parameter in the rest of the paper; its value is constrained to the range $0\leq\alpha\leq1$.

Setting $\varphi=\varphi'$  as well as $t = t'$, we obtain that the Green's function for radially and polarly separated points on the subtracted black hole background is given by:
\begin{equation}\label{greenu}
G_4(r,\theta;r',\theta')  = \frac{1+\alpha^2}{128\pi^2m^2(\Pi_c+\alpha^2\Pi_s)}\int_{-\infty}^\infty du \frac{\zeta}{\left[\zeta^2-\frac{(1+\cos\gamma)}{2}\right]^{3/2}}
\end{equation}
where we have changed variables to $u=c_+ z$. It is also convenient to use a dimensionless radial coordinate with origin at the outer horizon:
\begin{equation}\label{xdef}
x \equiv \frac{r-r_+}{r_+-r_-}\,.
\end{equation}
Note that $x=-1$ corresponds to the inner horizon if there is one and to the singularity if there isn't. In this notation, the $\zeta$ and $\cos \gamma$ functions appearing in (\ref{greenu}) read:
\begin{equation}\label{lambdadef}
\zeta = \sqrt{1+x}\sqrt{1+x'}\cosh u - \sqrt{x}\sqrt{x'}\cosh (\lambda u) \,\quad\quad \lambda \equiv \frac{\alpha^2 \Pi_c+\Pi_s}{\alpha^2 \Pi_s+\Pi_c}
\end{equation}
\begin{equation}\label{cdef}
\cos\gamma = \cos\theta \cos (\theta')+\sin\theta \sin(\theta') \cos(2 c u)\,\quad\quad c\equiv \frac{\alpha(\Pi_c-\Pi_s)}{\Pi_c+\alpha^2\Pi_s}
\end{equation}
As noted above, this exact form of $\zeta$ holds only when we are in the exterior region with $x,x' >1$; the form for general $(x, x')$ is given in the Appendix.

Note that setting $\Pi_c=1$, $\Pi_s=0$, $\theta=\theta'$, $x=0$, $x'=\epsilon$ in (\ref{greenu}), after a change of variables $w=\sinh u$ we recover eq.~18 from \cite{Cvetic:2015cca}, which corresponds to the Green's function with radial separation at the horizon of subtracted Kerr. In the cited reference this was computed as a sum over modes solving the 4d Euclidean wave equation, with no reference to the 5d embedding manifold. This provides a nontrivial validity check for our dimensional reduction procedure. Since the modes used in \cite{Cvetic:2015cca} correspond to the ``Hartle-Hawking'' Green's function, which is thermal with temperature $T=\kappa_+/2\pi$ as seen by co-rotating observers, we conclude that the dimensional reduction procedure from the $AdS^3\times S^2$ vacuum Green's function results in the Euclidean thermal co-rotating vacuum of the four-dimensional black hole. The outer horizon's surface gravity thus related to the temperature is given by:
\begin{equation}
\kappa_+=\frac{1}{4m}\frac{1-\alpha^2}{\Pi_c-\alpha^2\Pi_s}
\end{equation}

To close this section, we remark that the integral in (\ref{greenz}) is ill-defined for certain values of the eight 4d coordinates. The cases in which this happens can be separated into two kinds. Firstly, when both $r$ and $r'$ are larger than $r_-$ (i.e., none of the points lies within the inner horizon) the integral goes over a lightcone singularity if the points in the underlying 5d manifold are timelike separated. The Green's function is perfectly well defined when the separation is spacelike, though, and the definition of $G_4$ for timelike separation is achieved by analytic continuation with the usual $i \epsilon$ prescription for the Feynman Green's function.\footnote{This is analogous to how the vacuum Green's Function in d-dimensional Minkowski space can be obtained integrating over the embedding dimension the Green's Function in (d+1)-dimensional Minkowski space, analytically continuing the integral when it includes a lightcone singularity.  } Since we will compute the vacuum polarization by coincidence limit from a spacelike separation, this need not concern us further. Secondly, when either point has $r<r_-$, the integral goes over a singularity whether the separation is timelike or spacelike. We interpret this as meaning that the quantum state under consideration cannot be meaningfully extended to this internal region. This is in accordance with the situation for the Green's function for the BTZ black hole in three dimensions \cite{Steif:1993zv} as well as with the well-known instability of the inner horizon under perturbations in general 4d black holes \cite{Matzner:1979zz, Anderson:1993ni}.

\section{Evaluation of $\langle\phi^2\rangle$}

We will use angular point splittng to compute the vacuum polarization. Setting $x=x'$ in (\ref{greenu}), using trigonometric identities the Green's function can be recast as:
\begin{equation}\label{intwitpref}
G_4(\theta,\theta+\epsilon)  = \frac{1+\alpha^2}{64\pi^2m^2(\Pi_c+\alpha^2\Pi_s)} I_\epsilon
\end{equation}
\begin{equation}\label{Iepsdef}
 I_\epsilon = \int_0^\infty du \frac{\zeta(u)}{\left[\Delta_\epsilon(u)\right]^{3/2}}
\end{equation}
\begin{equation}\label{zetau}
\zeta(u)=(1+x)\cosh u -x \cosh(\lambda u)
\end{equation}
\begin{align}\label{Deltaepsu}
\Delta_\epsilon(u)=&\zeta^2(u)-1+\sin^2\theta \sin^2(c u/2)+\sin^2(\epsilon/2)\nonumber\\
&+(\cos\theta\sin\epsilon-2\sin\theta\sin^2(\epsilon/2))\sin^2(c u)
\end{align}
Even though derived here from the expression in the exterior region, the form of $\zeta$ given in (\ref{zetau}) is valid now for any value of $x$, as shown in the Appendix. Equation (\ref{intwitpref}) therefore gives the Green's function for polar point-splitting in the whole spacetime, although the result is ill-defined in the inner horizon internal region $x<-1$ as discussed above.

Our general strategy for evaluating the $\epsilon\to 0$ limit explicitly is analogous to the one deployed in \cite{Cvetic:2015cca}. We first split the integral in two subintervals, $(0,\eta)$ and $(\eta,+\infty)$, with $\eta=\epsilon^{1/3}$. In the lower interval, the integrand is expanded in a way such that the integral can be evaluated analytically with a controlled error. In the upper one, $\epsilon$ can be set to zero in the integrand without affecting the final result. The divergences are thus isolated and cancelled with appropriate counterterms coming from the Hadamard expansion of the two-point function, leaving an explicit analytic expression for the vacuum polarization in the coincidence limit.

Considering first the upper interval, let us call $G_\epsilon(u)$ the integrand in (\ref{Iepsdef}), and let $G_0(u) = G_{\epsilon=0}(u)$ and $\Delta_0(u) = \Delta_{\epsilon=0}(u)$. We claim that:
\begin{equation}\label{lowerwitheta}
I_\epsilon^>\equiv\int_\eta^\infty du\, G_\epsilon(u)\sim\int_\eta^\infty du \,G_0(u) 
\end{equation} 
where $\sim$ stands for equality up to terms vanishing in the limit $\epsilon\to 0$.
The reason is that the error involved in this replacement can be written as:
\begin{equation}
\int_\eta^\infty du \frac{\zeta(u)}{[\Delta_0(u)]^{3/2}}\left[1-\frac{1}{\left[1+\frac{\sin^2(\epsilon/2)+(\cos\theta\sin\epsilon-2\sin\theta\sin^2(\epsilon/2))\sin^2(c u)}{\Delta_0(u)}\right]^{3/2}}\right]
\end{equation}
and since $\Delta_0(u)=O(u^2)$ at small $u$ and is divergent at large $u$, it follows that the term having it as denominator is bounded by a constant of order $\epsilon$, and thus the integral is of order $\epsilon^{1/3}$.

The last integral in (\ref{lowerwitheta}) is still divergent as $\eta\to 0$, but the divergence is easily isolated explicitly  by adding and subtracting the leading terms in the expansion of $G_0$,  evaluating explicitly the added terms, and taking $\eta\to 0$ in the subtraction. This leads to: 

 \begin{align}\label{upper}
 I_\epsilon^>  &
\sim\frac{1}{2\epsilon^{2/3}(1+c^2v^2+x-\lambda^2 x)^{3/2}}\nonumber\\
 &-\frac{4c^4v^2+(\lambda^2-1)^2x(1+x)+4c^2v^2(1+x-\lambda^2 x)\ln \epsilon}{24(1+c^2v^2+x-\lambda^2 x)^{5/2}}\nonumber\\ 
& + \int_0^\infty du\Bigg[ \frac{\zeta(u)}{[\Delta(u)]^{3/2}}- \Bigg( \frac{1}{u^3(1+c^2v^2+x-\lambda^2 x)^{3/2}}\nonumber\\
 &+\frac{4c^4v^2+(\lambda^2-1)^2x(1+x)+4c^2v^2(1+x-\lambda^2 x)}{8u(1+u)(1+c^2v^2+x-\lambda^2 x)^{5/2}} \Bigg) \Bigg]
 \end{align}
where for compactness we introduce the notation $v=\sin\theta$. The upper interval's contribution is therefore one term divergent as $\epsilon^{-2/3}$ and two finite terms (one of which is expressed as an integral).

In the lower interval, we expand the numerator and denominator in the integrand:
\begin{equation}
I_\epsilon^< \equiv\int_0^\eta d u \frac{{\zeta}}{{\Delta}_\epsilon^{3/2}}\sim\int_0^\eta d u \frac{\tilde{\zeta}}{\tilde{\Delta}_\epsilon^{3/2}}\,,
\end{equation}
where $\tilde{\zeta}$ is the expansion of $\zeta$ to the second order in $u$ around $u=0$, and $\tilde{\Delta}_\epsilon$ the expansion of the $\Delta_\epsilon(u)$ to the fourth order in $u$ around $u=0$. The error involved can be shown to vanish in the limit $\epsilon\to 0$. 
The integrand now being a combination of terms of the form $\left((u^2+A^2)(u^2+B^2)\right)^{-3/2}$ and $u^2\left((u^2+A^2)(u^2+B^2)\right)^{-3/2}$,  the integral can be expressed in terms of the elliptic functions  $E(y,k)$ and $F(y,k)$ where $y =\arctan(\eta/A)$ and $k=1-\frac{A^2}{B^2}$. This result  is then  expanded for small $\epsilon$ using formulas from \cite{gustafson}, giving: 
\begin{align}\label{lower}
I_\epsilon^<&\sim -\frac{4}{\epsilon^2\sqrt{1+c^2v^2+x(1-\lambda^2)}}-\frac{2c^2v\sqrt{1-v^2}}{\epsilon(1+c^2v^2+x-\lambda^2x)^{3/2}}\nonumber\\
&-\frac{1}{2\epsilon^{2/3}(1+c^2v^2+x-\lambda^2x)^{3/2}}\nonumber\\
&+\frac{\left(-4c^4v^2-(\lambda^2-1)^2x(1-x)+4c^2v^2((\lambda^2-1)x-1)\right)\ln \epsilon}{12(1+c^2v^2+x-\lambda^2x)^{5/2}}\nonumber\\
&+\frac{1}{48(1+c^2v^2+x-\lambda^2x)^{5/2}}\times \nonumber\\
&
\left[(\lambda^2-1)x(-7-3x+\lambda^2(3x-1)\ln(1+c^2v^2+x(1-\lambda^2)))\right.\nonumber\\
&\left.+4(6-2c^4v^2(-5+v^2) 
+14c^2v^2(1+x-\lambda^2 x))\right.\nonumber\\
&\left.+x(10+x(4+\ln 8)+\lambda^4(-2-\ln 2+x(4+\ln 8))-\lambda^2(1+x)(8+\ln 64)+\ln 128 ) \right.\nonumber\\
&\left.+((\lambda^4-1)x+3c^2v^2(1+c^2+x-\lambda^2x))\ln(16(1+c^2v^2+x-\lambda^2x))\right]
\end{align}
The accuracy of this expression in the $\epsilon\to 0$ limit can be verified numerically. Note that the $\epsilon^{-2/3}$ divergence cancels with that of (\ref{upper}), and we are left with quadratic, linear and logarithmic divergences. These are cancelled subtracting from the Green's function the Hadamard expansion \cite{Christensen:1976vb}:
\begin{equation}\label{countersigma}
G_{div} = \frac{1+\frac{1}{12}R_{\mu\nu}\sigma^{,\mu}\sigma^{,\nu}}{8\pi^2\sigma} -\frac{1}{96\pi^2} R \ln (\mu^2 \sigma)\,,
\end{equation}
where $\mu$ is an arbitrary mass scale. We express the halved squared geodesic distance $\sigma$ in terms of the coordinate separation $\Delta X^\mu=x^\mu-x'^\mu$ (which in our case is $-\epsilon\delta^\mu_\theta$) using the expansion:
\begin{equation}
\sigma=\frac{1}{2}g_{\mu\nu}\Delta x^\mu\Delta x^\nu + A_{\mu\nu\rho}\Delta x^\mu\Delta x^\nu\Delta x^\rho
+ B_{\mu\nu\rho\lambda}\Delta x^\mu\Delta x^\nu\Delta x^\rho\Delta x^\lambda+\cdots
\end{equation}
Here $A_{\mu\nu\rho}$ and $B_{\mu\nu\rho\lambda}$ have expressions terms of derivatives of the metric, provided explicitly in \cite{Ottewill:2008uu}; see also \cite{Synge}. Writing $G_{div}$ in terms of $\epsilon$ gives linear, quadratic and logarithmic divergences that match exactly those of the sum of (\ref{upper}) and (\ref{lower}) (including the prefactor in (\ref{intwitpref})). There is also an additional finite piece coming from terms of $G_{div}$ that are $O(1)$ in $\epsilon$. 
Combining all the pieces the full expression for the vacuum polarization is:

\begin{align}\label{general}
\langle\phi^2&\rangle  = \lim_{\epsilon\to 0}\left(G_4(\theta,\theta+\epsilon)-G_{div}(\theta,\theta+\epsilon)\right) \,=\,\frac{1+\alpha^2}{64\pi^2m^2(\Pi_c+\alpha^2\Pi_s)}\times\nonumber\\
&\Bigg\{\frac{1}{48(1+c^2v^2+x-\lambda^2x)^{5/2}}\,\Bigg[
4\Big(6-2(-5+4\lambda^2+\lambda^4)x-2c^4v^2(-5+v^2-6\ln 2)\nonumber\\
&+(\lambda^2-1)^2x(\ln 8+x(4+\ln 8))-2c^2v^2(-1+(\lambda^2-1)x)(7+\ln 64)\Big)\nonumber\\
&+3\Big(4c^4v^2+(\lambda^2-1)^2
x(1+x)+4c^2v^2(1+x-\lambda^2 x)\Big)\nonumber\\
&\ln(1+c^2v^2+x-\lambda^2x)\Bigg]
+\left(4c^2+(\lambda-1)(\lambda-1+\sqrt{4c^2+(\lambda-1)^2})\right)\nonumber\\
&\times\frac{2c^4+(\lambda-1)^3(\lambda-1+\sqrt{4c^2+(\lambda-1)^2})+2c^2(\lambda-1)(2\lambda-2+\sqrt{4c^2+(\lambda-1)^2})}{24\sqrt{2}\sqrt{4c^2+(\lambda-1)^2}\left(2c^2+(\lambda-1)(\lambda-1+\sqrt{4c^2+(\lambda-1)^2})\right)\left(1+c^2v^2+x-\lambda^2x\right)^{5/2}}
\nonumber\\
&\times\Bigg[
4\Big(c^2(4-(21+19c^2+\lambda^2)v^2+5c^2v^4)-3-\lambda^2\Big)\nonumber\\
&-(\lambda^2-1)x\left(\lambda^2-25+c^2(16-80v^2)\right)
-13(\lambda^2-1)^2 x^2\Bigg]\nonumber\\
&+\int_0^\infty du\Bigg[ \frac{\zeta(u)}{[\Delta(u)]^{3/2}}- \Bigg( \frac{1}{u^3(1+c^2v^2+x-\lambda^2 x)^{3/2}}\nonumber\\
&+\frac{4c^4v^2+(\lambda^2-1)^2x(1+x)+4c^2v^2(1+x-\lambda^2 x)}{8u(1+u)(1+c^2v^2+x-\lambda^2 x)^{5/2}} \Bigg) \Bigg]\Bigg\}\,.
\end{align}
 This is the main result of the paper, expressing the vacuum polarization of a quantum massless minimally coupled scalar field in a subtracted geometry black hole spacetime, any value of the mass, rotation and charge parameters, in terms of which $\alpha$, $\lambda$ and $c$ were defined above in (\ref{alphadef}), (\ref{lambdadef}) and (\ref{cdef}) respectively.  The expression is valid for any values of the angular coordinate $v$ and for all values of radial coordinate $x$ outside the inner horizon ($x\geq-1$). The result has several explicitly evaluated terms and one expressed as an integral with no closed form, but easily evaluated numerically. This expression is correct up to a term of the form $C R(x)$, where $C$ is an arbitrary number, and the Ricci scalar $R$ is $R(x)=\tilde{R}/32m^2$, with:
\begin{align}
\tilde{R}=&\frac{3(1+\alpha^2)^2(\Pi_c-\Pi_s)^2}{\left[-2\alpha^2\Pi_c\Pi_s(v^2-1)+\Pi_c^2(1+\alpha^2v^2+x-\alpha^4x)+\Pi_s^2(\alpha^2v^2-x+\alpha^4(1+x))\right]^{5/2}}\nonumber\\
&\times\Big[4\alpha^2(\pi_c^2+\alpha^2\Pi_s^2)v^2+(\alpha^2-1)x((\alpha^4-1)(\Pi_c+\Pi_s)^2+4\alpha^2v^2(\Pi_s^2-\Pi_c^2))\nonumber\\
&+(\alpha^2+1)(\alpha^2-1)^2(\Pi_c+\Pi_s)^2x^2\Big]\,.
\end{align}
In the next section we will compare with previously known results, as well as examining particular limiting values and discussing their physical significance.

\section{Discussion}

The first check on our result (\ref{general}) is whether it agrees with the results in \cite{Cvetic:2015cca} when evaluated at the outer horizon of a Kerr black hole. This implies setting $\lambda=\alpha^2$, $c=\alpha$ $\Pi_c=1$, $\Pi_s=0$, and $x=0$. Since the calculation in the cited reference was done by radial point splitting leading to a differently structured result, the comparison is not possible term by term but only between the total results. Numerical evaluation of the $u$-integral in each result shows that both results are indeed equal throughout the horizon, up to a multiple of the Ricci scalar $R$ (in other words, the difference between both results divided by $R$ is a $\theta$-independent constant).

A more direct comparison is available in the static case. Setting $c=\alpha=0$, $x=0$, and $\lambda=\Pi_s/\Pi_c$, the $u$-integral in (\ref{general}) becomes elementary and we obtain an explicit formula for the vacuum polarization at the outer horizon of general static black holes:
\begin{equation}
\langle\phi^2\rangle\Big|_{r_+,\alpha=0} = \frac{\Pi_c^2-\Pi_s^2}{768\pi^2m^2\Pi_c^3}\,,
\end{equation}
This matches the results obtained in \cite{Cvetic:2014eka}.

We are now in position to extend both these previous results to obtain the vacuum polarization at the horizon of a fully general subtracted black hole, with both rotation and charge parameters being nonzero. For simplicity we exhibit the result only in the subtracted Kerr-Newman case, where there is a single charge parameter $\delta$ and $\Pi_c=\cosh^4\delta,\,\Pi_s=\sinh^4\delta$. The result of evaluating (\ref{general}) in this case for $x=0$ is plotted in Figures 1 and 2 as a function of $v=\sin\theta$ for different combinations of the rotation and charge parameters.

\begin{figure}
\begin{center}
\includegraphics[width=0.7\textwidth]{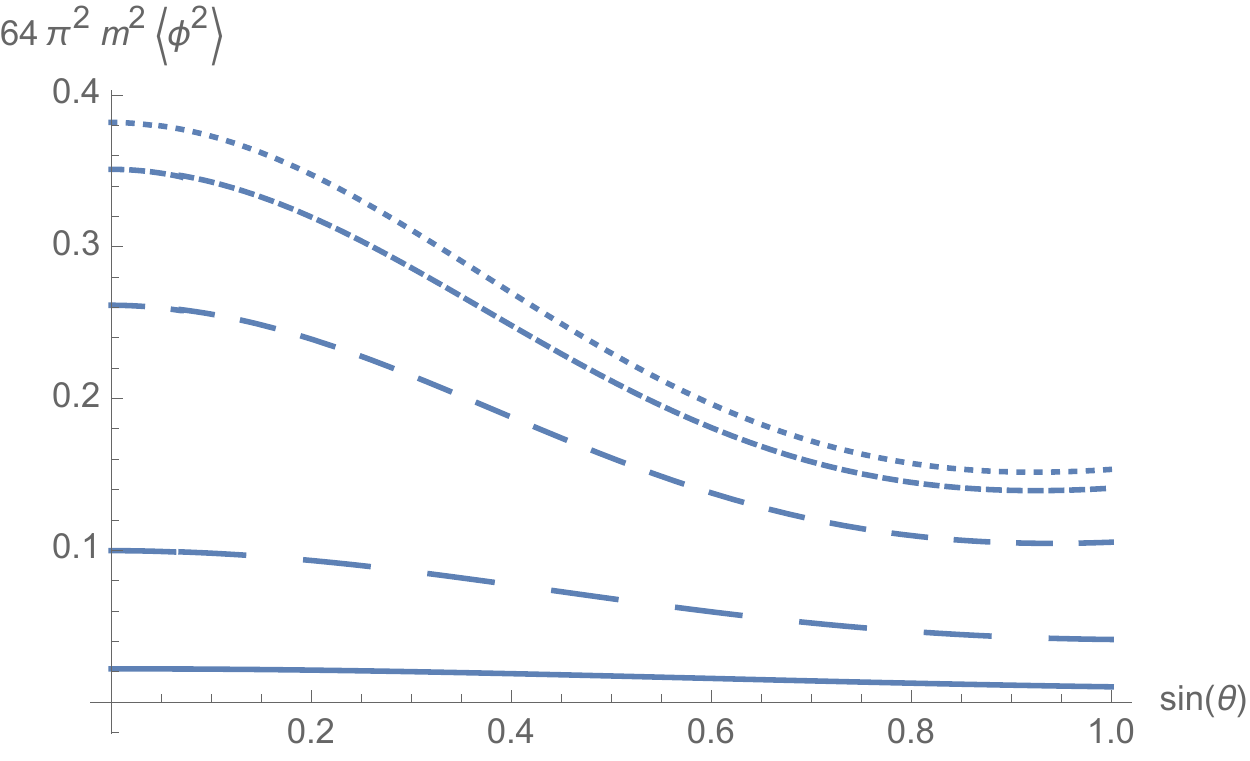}
\end{center}
\caption{$(64\pi^2m^2)\langle\phi^2\rangle$ at the Kerr-Newman horizon, for $a/r_+=\alpha=0.75$. The charge parameter $\delta$ takes the values 0, 0.2, 0.4, 0.7 and 1 for the dotted, small-dashed, medium-dashed, large-dashed and full lines respectively.}
\label{varyingcharge}
\end{figure}

\begin{figure}
\begin{center}
\includegraphics[width=0.7\textwidth]{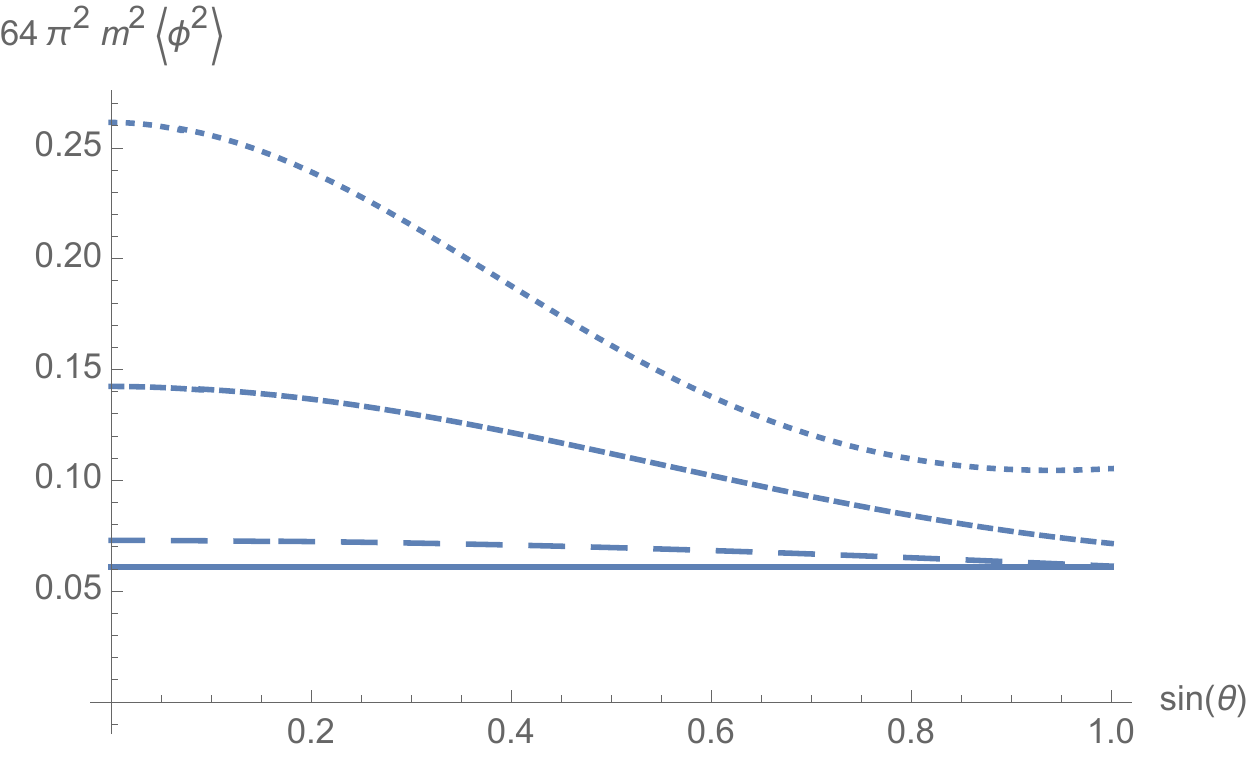}
\end{center}
\caption{$(64\pi^2m^2)\langle\phi^2\rangle$ at the Kerr-Newman horizon, for $\delta=0.4$ (corresponding to $Q/M=0.66$). The rotation parameter $\alpha=a/r_+$ takes the values 0.75, 0.4, 0.2, and 0 for the dotted, small-dashed, medium-dashed and full lines respectively.}
\label{varyingrotation}
\end{figure}

It can be seen in Figure 1 that increasing the charge at fixed angular momentum lowers the vacuum polarization, making it vanish in the limit $\delta\to\infty$. In Figure 2 it is seen that lowering the angular momentum at fixed charge flattens the angular profile as we approach the spherically symmetric static limit. Though the plots are presented for a particular value of the arbitrary constant multiplying $R$, these qualitative features are independent of it.

Let us now consider the vacuum polarization beyond the horizon. In the static case with at least one vanishing charge $\delta_I$, in which $\Pi_s=0$ and $a=0$, we can obtain a simple closed-form expression for the vacuum polarization as a function of $x$. This is easiest done not from our expression (\ref{general}) but taking a step back to evaluate the $u$-integral in (\ref{lowerwitheta}). The parameters $\alpha,c,\lambda$ are all 0, so we have:
\begin{equation}
G_0(u)=\frac{\zeta(u)}{[\Delta_0(u)]^{3/2}}=\frac{(1+x)\cosh u - x}{\left[((1+x)\cosh u - x)^2-1\right]^{3/2}}
\end{equation}
Then (\ref{lowerwitheta}) can be computed in closed form and, after joining with the explicitly computed terms from (\ref{lower}) and (\ref{countersigma}), we obtain:
\begin{equation}\label{schwa}
\langle\phi^2\rangle\Big|_{\alpha=\Pi_s=0}=\frac{1}{768\pi^2m^2\Pi_c}\frac{-12-8x+6(2+x)\sqrt{1+x}+3x^2\left(-2+\ln\left(\frac{256(1+x)}{2+x+2\sqrt{1+x}}\right)\right)}{4x(1+x)^{3/2}}
\end{equation}
This does not include the arbitrary $R$ term, which takes the form of an arbitrary constant times
\begin{equation}
R\Big|_{\alpha=\Pi_s=0}=\frac{3x}{32m^2\Pi_c(1+x)^{3/2}}
\end{equation}
Asymptotically at large $x$, both $R$ and the other contributions to $\langle\phi^2\rangle$ behave as $x^{-1/2}$. Hence the form of the decay is universal but the constant in front of it is not. At the outer horizon, $R=0$ and $\langle\phi^2\rangle=(768\pi^2m^2\Pi_c)^{-1}$.

The singularity is approached as $x\to-1$. In this limit, the singularity the vacuum polarization diverges as:
\begin{equation}\label{sing}
\langle\phi^2\rangle\Big|_{x\to-1}\sim \frac{1}{768\pi^2m^2\Pi_c}\left(\frac{C-3\ln(1+x)}{(1+x)^{3/2}}+O\left(1+x\right)^{-1}\right)
\end{equation}
where $C$ is an arbitrary number. 

Figure 3 exhibits the $x$-dependence of $\langle\phi^2\rangle$ for Schwarzschild (and other $\Pi_s=0$ static black holes), as given above in (\ref{schwa}), with the plot scaled to have the horizon value 1.
\begin{figure}
\begin{center}
\includegraphics[width=0.7\textwidth]{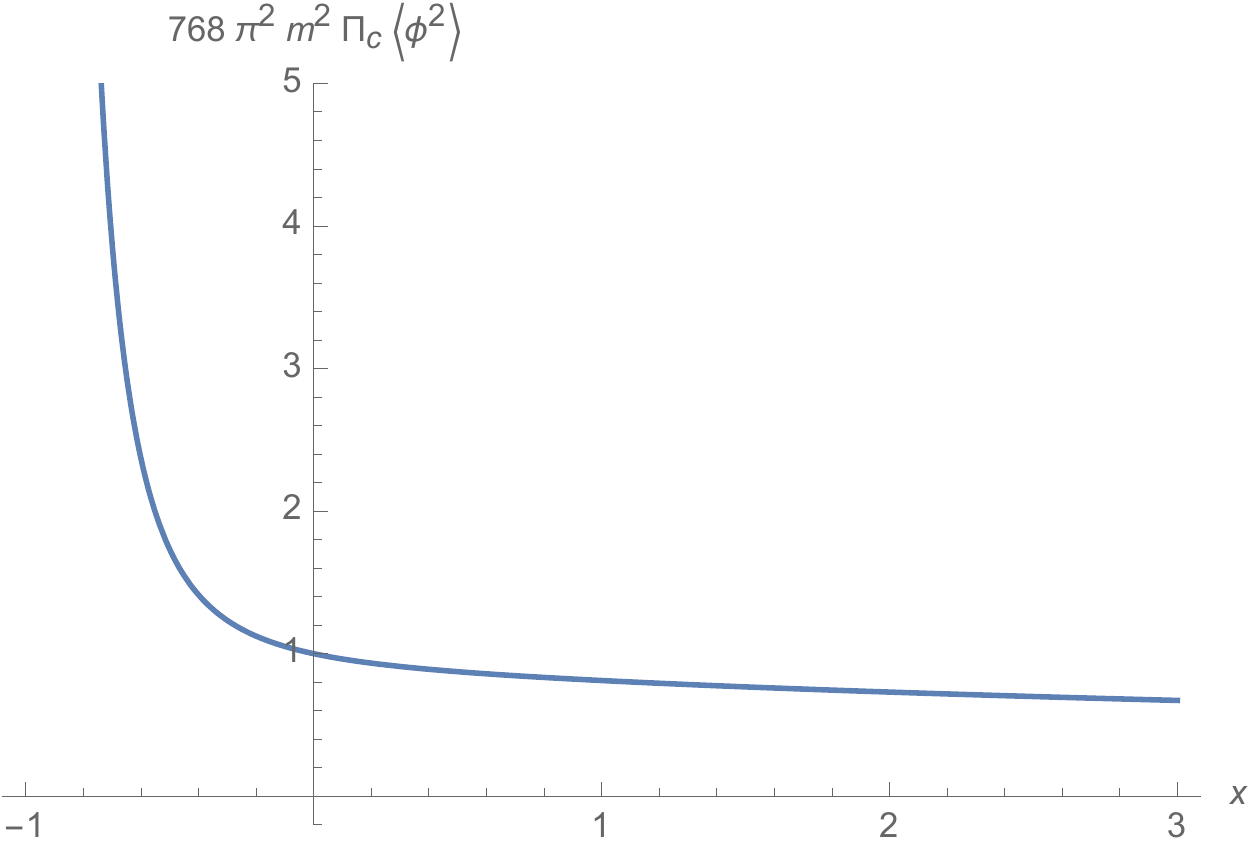}
\end{center}
\caption{$(768\pi^2m^2\Pi_c)\langle\phi^2\rangle$ as a function of $x$ for $\alpha=0=\Pi_s$. The constant multiplying $R(x)$ in the arbitrary term added to (\ref{schwa}) is set as $-\pi^2m^2\Pi_c$.}
\label{schwarana}
\end{figure}
Note that this plot corresponds to a particular value of the added $R$-term, and that only the behaviors at the singularity and at infinity are physical (as well as the horizon value, since $R(x=0)=0$). However, note as well that the behavior near the singularity is universal since the $C$-term is subdominant in (\ref{sing}). The vacuum polarization diverges at the spacelike singularity of Schwarzschild (and other static black holes with at least one vanishing charge) with the leading divergence being $-\ln(y)/256\pi^2m^2\Pi_c y^{3/2}$, where $y=1+x$ is the dimensionless radial coordinate (timelike in the inner region) translated to vanish at the singularity.

It is natural to inquire about a comparison between the vacuum polarizations beyond the horizon for subtracted Schwarzschild and for standard Schwarzschild. A simple analytical approximation to the latter (in the Hawking-Hartle state) was developed by Page and Whiting \cite{Page:1982fm} and reads
\begin{equation}
\langle\phi^2\rangle_{PW}^{\mathrm{Sch}}=\frac{1}{768\pi^2m^2}\left(1+\frac{2m}{r}+\frac{4m^2}{r^2}+\frac{8m^3}{r^3}\right)
\end{equation}
This approximation is reliable up very close to the singularity \cite{Candelas:1985ip}.  It is seen that $\langle\phi^2\rangle$ in standard Schwarzschild goes to a constant asymptotic value (characteristic of thermal radiation) very far from the black hole, while in the subtracted case it vanishes.\footnote{A heuristic explanation is that the order of magnitude of the Page-Whiting approximation is given by the $\kappa^2\Omega^{-2}$ where $\kappa$ is the surface gravity and $\Omega^2$ the conformal factor relating the metric to an ultrastatic metric with $g_{00}=-1$. Standard Schwarzschild is asymptotically flat, but subtracted Schwarzschild is not and in it $\Omega^{-2}$ vanishes asymptotically. However, in any case the Page-Whiting approximation is derived for Einstein spaces and conformally coupled fields, so there is no reason to expect it to be accurate in our case.} As for the divergence approaching the singularity, insofar as the Page-Whiting approximation provides the right order of magnitude it is seen that the divergence is stronger ($\sim y^{-3}$) in standard Schwarzschild than in subtracted Schwarzschild.

\begin{figure}[ht]
\begin{center}
\includegraphics[width=0.7\textwidth]{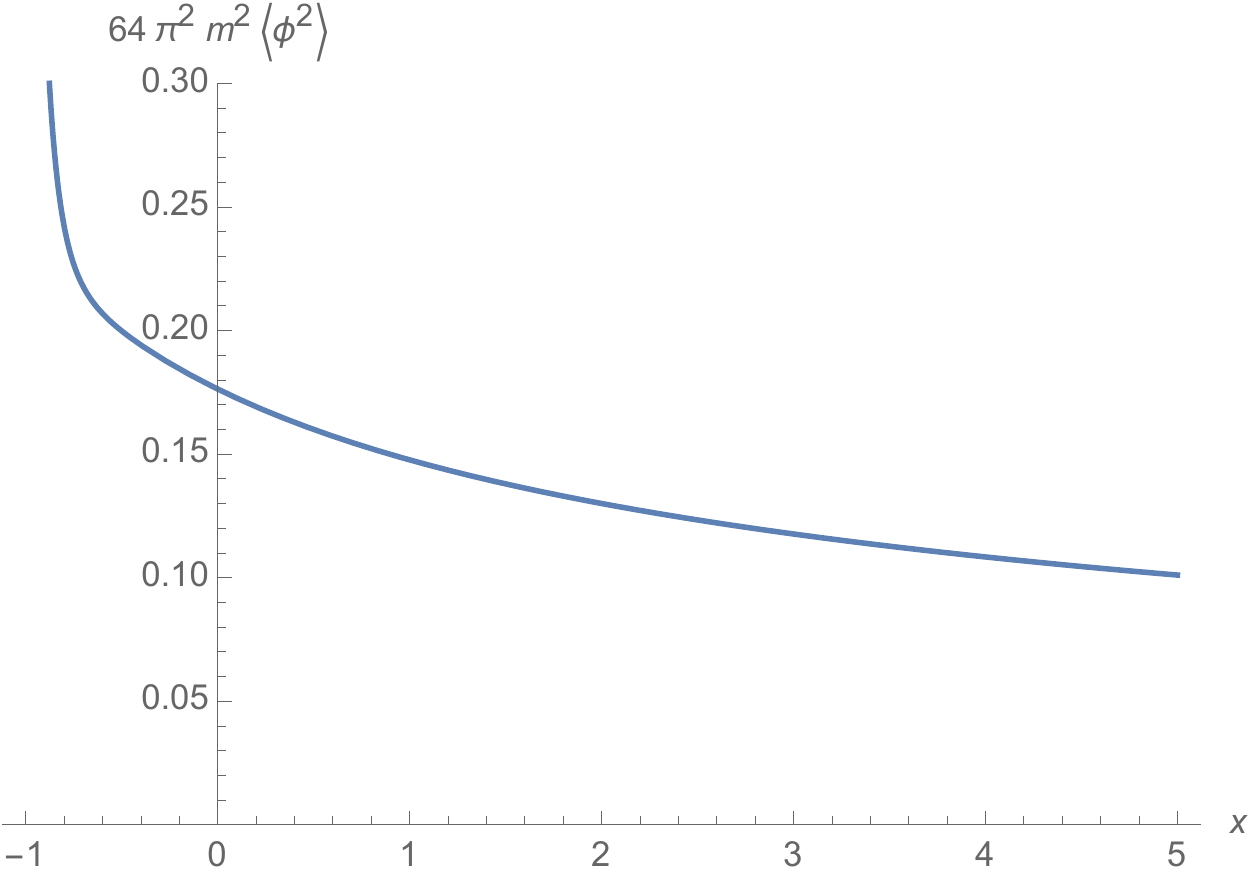}
\end{center}
\caption{$(64\pi^2m^2)\langle\phi^2\rangle$ at the equatorial plane as a function of the dimensionless radial coordinate $x$ for the subtracted Kerr black hole with $\alpha=1/2$. The divergence at $x=-1$ corresponds to the inner horizon.}
\label{kerrxplot}
\end{figure}

For black holes with two horizons, the vacuum polarization is well-defined only up to the inner horizon. As an example, the result for the subtracted Kerr black hole with $\alpha=1/2$ is plotted in Fig. \ref{kerrxplot} (at the equatorial plane $\theta=\pi/2$). Note that the vacuum polarization diverges as the inner horizon ($x=-1$) is approached. Other cases with two horizons, such as subtracted Kerr-Newman, behave in a qualitatively similar way. 

\section{Conclusions}

The main result of this paper, given in equation (\ref{general}), is an analytical expression for the vacuum polarization of a massless, minimally coupled scalar field in the general subtracted four-dimensional black hole background. The remarkable features of the result are its validity throughout the spacetime from the inner horizon to the asymptotic boundary, and its validity for black holes with arbitrary charged and rotation parameters. To our knowledge, this is the first such expression ever obtained in four-dimensional black holes. 

We have shown that the result correctly recovers previous evaluations at the subtracted static and Kerr horizons \cite{Cvetic:2014eka,Cvetic:2015cca} and reduces to a simple closed-form expression (\ref{schwa}) valid in the subtracted Schwarzschild case. In this case the vacuum polarization diverges at the singularity. In the general case, it diverges at the inner horizon, and the Green's function characterizing the quantum state is ill-defined in the inner region.

We expect the techniques we used to evaluate $\langle\phi^2\rangle$ to be extendable to the computation of the stress-energy tensor $\langle T_{\mu\nu}\rangle$. If this computation is achieved, it could be used to study the backreaction of the quantum field upon the geometry in an analytical way. Among other questions, this would illuminate the quantum effects near the singularity. Moreover, our general expressions for the Green's function (provided in the Appendix) can also be used to probe quantum effects in this geometry in other ways, such as the response of a particle detector. 

\section*{ Acknowledgements} 

We would like to thank Gary Gibbons, Finn Larsen and Zain Saleem for discussions and collaborations on related topics. The research of M~C.~is supported in part by the DOE Grant Award de-sc0013528, the Fay R. and Eugene L. Langberg Endowed Chair and the Slovenian Research Agency (ARRS). The research of A.~S.~ is supported by the NSF grant PHY-150541.

  \appendix
\section{Full expression for the Green's Function in each region}

The expression for the Green's function on the 4-dimensional subtracted metric, as an integral of the five-dimensional Green's function on  $AdS^3\times S^2$, was provided above in (\ref{greenz}) and reads:
\begin{equation}\label{greenzapp}
G_4(t,r,\theta,\varphi;t'r',\theta',\varphi')  =\frac{1}{2\sqrt{2}\pi^2l^3}\int_{-\infty}^{+\infty}dz\frac{\zeta(z;0)}{[2\zeta^2(z;0)-1-\cos\gamma(z;0)]^{3/2}}
\end{equation}
where $\zeta$ and $\cos\gamma$ are given in (\ref{zetagamma})  (their dependence on the 4d coordinates is suppressed). The expression of $\cos\gamma$ in terms of the four-dimensional coordinates plus $z$ is always obtained replacing $\bar{\varphi}$  in it with the first line of (\ref{identifications}), but the expression of $\zeta$ is more complicated and differs depending on the values of $r,r'$. To express it in a succinct way it is convenient to define the auxiliary coordinates:
\begin{eqnarray}
T & = & \frac{R_+ t_3-l R_-\varphi_3}{l^2} \label{Tdef} \\
\Phi & = &  \frac{l R_+ \varphi_3 - R_- t_3}{l^2} \label{Phidef}
\end{eqnarray}
defined in terms of the 3d BTZ coordinates and the parameters $R_\pm$ defined in (\ref{identifications}). Using for the radial coordinate the dimensionless $x$ as defined in (\ref{xdef}), the $AdS^3$ distance function $\zeta$ takes the following form:
\begin{eqnarray}
\zeta_{11}&=&\sqrt{1+x}\sqrt{1+x'}\cosh(\Phi-\Phi')-\sqrt{x}\sqrt{x'}\cosh(T-T')\cr
\zeta_{22}&=&\sqrt{1+x}\sqrt{1+x'}\cosh(\Phi-\Phi')+\sqrt{-x}\sqrt{-x'}\cosh(T-T')\cr
\zeta_{33}&=&-\sqrt{-1-x}\sqrt{-1-x'}\cosh(\Phi-\Phi')+\sqrt{-x}\sqrt{-x'}\cosh(T-T')\cr
\zeta_{12}&=&\sqrt{1+x}\sqrt{1+x'}\cosh(\Phi-\Phi')-\sqrt{x}\sqrt{-x'}\sinh(T-T')\cr
\zeta_{13}&=&-\sqrt{1+x}\sqrt{-1-x'}\sinh(\Phi-\Phi')-\sqrt{x}\sqrt{-x'}\sinh(T-T')\cr
\zeta_{23}&=&-\sqrt{1+x}\sqrt{-1-x'}\sinh(\Phi-\Phi')+\sqrt{-x}\sqrt{-x'}\cosh(T-T')
\end{eqnarray}
Here the substcripts 1,2,3 label respectively the region external to the outer horizon ($x>0$), the middle region ($-1<x<0$) and the region internal to the inner horizon ($x<-1$), so that for example $\zeta_{12}$ is to be used for computing the Green's function when $x>0$ and $-1<x'<0$. Using the relations in (\ref{identifications}) together with (\ref{Tdef}, \ref{Phidef}), this is enough to express fully the Green's function in terms of four-dimensional coordinates, the subtracted black hole parameters, and an integrated-upon variable $z$. As mentioned at the end of Section 2.2, the $z$-integral has to be found through analytic continuation when the points are in the middle/outer regions and timelike separated; on the other hand, it is completely undefined when one of the points is in the inner region $x<-1$, owing to the failure of the quantum state to be defined in this region. It is easily checked that with only $\theta$-separation the Green's function is given by (\ref{intwitpref}-\ref{Deltaepsu}) in both the external and the middle region, which validates our computation of the vacuum polarization in Section 3.

\end{document}